\documentclass[supp]{new_tlp} 

\usepackage[english]{babel}

\usepackage{graphicx}
\usepackage{paralist}

\usepackage{ifthen}
\usepackage{url}
\usepackage{amssymb}
\usepackage{amsmath}
\usepackage{import}
\usepackage{xparse}
\usepackage{amsmath}
\renewcommand*\ttdefault{cmvtt}
\usepackage[usenames,dvipsnames]{xcolor}
\usepackage{xspace}
\usepackage{datatool}
\usepackage{glossaries}

\newcommand{\m}[1]{\ensuremath{#1}\xspace}
\newcommand{\trval}[1]{\m{\mbox{\bf #1}}}

	\newcommand{\limplies}{\m{\Rightarrow}}

	\newcommand{\lrule}{\m{\leftarrow}}
	\newcommand{\cause}{\m{\stackrel{c}{\lrule}}}
	
	\newcommand{\ltrue}{\trval{t}}

	\newcommand{\Tr}{\ltrue}

	\newcommand{\voc}{\m{\Sigma}}

	\newcommand{\struct}{\m{I}}
	
	\newcommand{\I}{\m{\mathcal{I}}}
	
	\newcommand{\J}{\m{\mathcal{J}}}

	\newcommand{\D}{\m{\Delta}}

	\newcommand{\set}{\m{S}}
	\NewDocumentCommand\inter{g+g}{%
	  \IfNoValueTF{#1}
	    {\struct}
	    {\m{#1^{#2}}}}

	\newcommand{\xxx}{\m{\overline{x}}}

	\newcommand{\ddd}{\m{\overline{d}}}

	\newcommand{\ttt}{\m{\overline{t}}}

	\renewcommand{\int}{\m{\mathbb{Z}}}

	\NewDocumentCommand\subs{g+g}{%
	  \IfNoValueTF{#1}
	    {\m{/}}
	    {\m{#1/ #2}}}

	\newcommand{\logicname}[1]{\text{\sc #1}\xspace}

	\newcommand{\fodot}{\logicname{FO(\ensuremath{\cdot})}}
	
	\newcommand{\foid}{\logicname{FO(\ensuremath{ID})}}
	
	\newcommand{\foidplus}{\logicname{C-Log}} 
	\newcommand{\clog}{\logicname{C-Log}}
	\newcommand{\foclog}{\logicname{FO(C)}}

	\newcommand{\fo}{\FO}
	
	\newcommand{\cpl}{\logicname{CP}-logic\xspace}

\newcommand{\ouracronym}[3]{%
	\newacronym{#1}{#2}{#3}
	\expandafter\newcommand\csname #1\endcsname{\gls{#1}\xspace}%
}
	\ouracronym{FO}{FO}{first-order logic}
	\ouracronym{MX}{MX}{Model Expansion}
	\ouracronym{MO}{MO}{Model Optimization}
	\ouracronym{ASP}{ASP}{Answer Set Programming}
	\ouracronym{TP}{TP}{Theorem Proving}
	\ouracronym{LP}{LP}{Logic Programming}
	\ouracronym{CP}{CP}{Constraint Programming}
	\ouracronym{KR}{KR}{Knowledge Representation}
	\ouracronym{CSP}{CSP}{Constraint Satisfaction Problem}
	\ouracronym{SMT}{SMT}{SAT Modulo Theories}
	\ouracronym{KBS}{KBS}{knowledge-base system}
	\ouracronym{NNF}{NNF}{Negation Normal Form}
	\ouracronym{FNNF}{FNNF}{Flat Negation Normal Form}
	\ouracronym{DefNNF}{DefNNF}{Definition Negation Normal Form}
	\ouracronym{CDCL}{CDCL}{Conflict-Driven Clause-Learning}
	\ouracronym{LCG}{LCG}{Lazy Clause Generation}

	\makeatletter
	\def\ifenv#1{
	\def\@tempa{#1}%
	\def\@ttempa{#1*}%
	\ifx\@tempa\@currenvir
	\expandafter\@firstoftwo
	\else
	\expandafter\@secondoftwo
	\fi
	}
	\makeatother

	\newcommand{\ddrule}[4]{\ensuremath{#1 \leftarrow #2 & \{#3\} & #4}}
	\newcommand{\drule}[2]{\ensuremath{#1 & \leftarrow & #2}}

	\newcommand{\darule}[4]{\ensuremath{#1 \leftarrow #2 & \{#3\} & #4}}
	\newcommand{\arule}[2]{\ensuremath{#1 \, &\leftarrow \, #2}}

	\newcommand{\LNDRule}[2]{
	\ifenv{array}
	{\drule{#1}{#2}}
	{ \ifenv{align}
		{\arule{#1}{#2}}
		{\ifenv{align*}
		{\arule{#1}{#2}}
		{ERROR: using LDRule in unsupported environment: \@currenvir}
		}
	}
	}

	\newcommand{\LDRule}[4]{
	\ifenv{array}
	{\ddrule{#1}{#2}{#3}{#4}}
	{ \ifenv{align}
		{\darule{#1}{#2}{#3}{#4}}
		{\ifenv{align*}
		{\darule{#1}{#2}{#3}{#4}}
		{ERROR: using LDRule in unsupported environment: \@currenvir}
		}
	}
	}

	\NewDocumentCommand\LRule{m+g+g+g}{%
		\IfNoValueTF{#2}%
		{#1.&}{%
		\IfNoValueTF{#3}
		{\LNDRule{#1}{#2.}}
		{\LDRule{#1}{#2.}{#3}{#4}}%
		}
	}

	\NewDocumentCommand\CLRule{m+g}{%
	\ifenv{array}
	{\cdrule{#1}{#2}}
	{ \ifenv{align}
		{\carule{#1}{#2}}
		{\ifenv{align*}
			{\carule{#1}{#2}}
			{ERROR: using CLRule in unsupported environment: \@currenvir}
		}
	}
	}

	\NewDocumentCommand\carule{m+g}{%
		\IfNoValueTF{#2}
			{\ensuremath{#1.}}
			{\ensuremath{#1 \, &\cause \, #2}}}
	\NewDocumentCommand\cdrule{m+g}{%
		\IfNoValueTF{#2}
			{\ensuremath{#1.}}
			{\ensuremath{#1 & \cause & #2}}}

	\newcommand{\algrule}[4]{
	\hbox{{#1}:}& 
	\quad #2 ~\longrightarrow~ #3 
	\hbox{~ if } #4\\
	}

	\newcommand{\AlgoRule}[4]{
	\ifenv{array}
	{\algrule{#1}{#2}{#3}{#4}}
		{ERROR: using AlgoRule in unsupported environment: \@currenvir}
	}

\newcommand{\commentstyle}{\color{Gray}}

	\usepackage[final]{listings}

	\lstdefinelanguage{idp}{
		morekeywords=[1]{namespace,vocabulary,theory,structure,procedure,term,set,formula, spec, specification},
		morekeywords=[2]{include,using,type,isa,contains,partial,extern,LFD,GFD,constructed,from,constraint,func,pred,supertype,of,subtype,define},
		morekeywords=[3]{int,float,char,string,nat},
		morekeywords=[4]{if,then,else,for,end},
		morecomment=[s]{/*}{*/},	
		morecomment=[l]{//}
	}
	\newcommand{\ignore}[1]{}

	\newboolean{nocomments}
	\setboolean{nocomments}{false}

	\newboolean{commentmargin}
	\setboolean{commentmargin}{true}

	\newcommand{\namedcomment}[3]{
		\ifthenelse{\boolean{nocomments}}
		{} 
		{ 
			\ifthenelse{\boolean{commentmargin}}
				{ {\color{#3} \marginpar{\color{#3}\sc #2}#1}  } 
				{  {\color{#3} {\sc #2}: #1}  } 
		}
	}
	\newcommand{\mnamedcomment}[3]{\ifthenelse{\boolean{nocomments}}{}{{\marginpar{ \color{#3}{\sc #2}:#1}}}}

	\usepackage{soul}

\newcommand{\keyword}[2]{%
	\expandafter\newcommand\csname #1\endcsname{#2\xspace}%
	\expandafter\newcommand\csname #1s\endcsname{#2s\xspace}%
	\expandafter\newcommand\csname #1ness\endcsname{#2ness\xspace}%
}

\usepackage{etoolbox}

\newcommand\setcitation[2]{%
  \csdef{mycommoncitation#1}{#2}}
\newcommand\getcitation[1]{%
  \csuse{mycommoncitation#1}}

\setcitation{IDP}{WarrenBook/DeCatBBD14}
\setcitation{fodot}{tocl/DeneckerT08}
\setcitation{CP}{Apt03}
\setcitation{EZCSP}{lpnmr/Balduccini11}
\setcitation{KR}{Baral:2003}
\setcitation{ASPComp2}{lpnmr/DeneckerVBGT09}
\setcitation{ASPComp3}{LPNRM/Calimeri11}
\setcitation{ASPComp4}{conf/lpnmr/AlvianoCCDDIKKOPPRRSSSWX13}
\setcitation{CPSupport}{ictai/DeCat13}
\setcitation{functionDetection}{iclp/DeCatB13}
\setcitation{fodot2asp}{Denecker12}
\setcitation{Inca}{aaai/DrescherW11}
\setcitation{csp2asp}{ijcai/DrescherW11}
\setcitation{DPLLT}{cav/GanzingerHNOT04}
\setcitation{AspInPractice}{synthesis/2012Gebser}
\setcitation{clasp}{ai/GebserKS12}
\setcitation{oclingo}{kr/GebserGKOSS12}
\setcitation{gringo}{lpnmr/GebserST07}
\setcitation{cmodels}{aaai/GiunchigliaLM04}
\setcitation{inputster}{tplp/Jansen13}
\setcitation{DLV}{tocl/LeonePFEGPS06}
\setcitation{LearningPaper}{corr/Blockeel13} 
\setcitation{clog}{iclp/BogaertsVDV14} 
\setcitation{inferenceClog}{ecai/BogaertsVDV14}
\setcitation{examplesClog}{nmr/BogaertsVDV14b} 
\setcitation{AFT}{DeneckerBV12}
\setcitation{KBS}{iclp/DeneckerV08}
\setcitation{KBPE}{inap/DePooterWD11}
\setcitation{lazyGrounding}{corr/CatDSB14} 
\setcitation{}{}
\setcitation{}{}
\setcitation{}{}
\setcitation{}{}
\setcitation{}{}
\setcitation{}{}
\setcitation{}{}
  
\newcommand\mycite[1]{%
      \ifcsname mycommoncitation#1\endcsname%
   \cite{\getcitation{#1}}%
  \else%
    \cite{#1}
  \fi%
}

\newcommand{\univBQ}[2]{\m{\forall #1 [#2]}}
\newcommand{\exisBQ}[2]{\m{\exists #1 [#2]}}

\newcommand{\pred}[1]{\m{\mathrm{#1}}}

\usepackage{amsmath}
\NewDocumentCommand\lattice{g+g}{%
  \IfNoValueTF{#1}
    {\m{L}}
    {\m{\left(#1,#2\right)}}%
}

\newcommand{\Bilat}[1]{\m{#1^2}}
\NewDocumentCommand\bilat{g}{%
  \IfNoValueTF{#1}
    {\Bilat{\lattice}}
    {\Bilat{#1}}%
}

\newcommand{\cBilat}[1]{\m{#1^c}}
\NewDocumentCommand\cbilat{g}{%
  \IfNoValueTF{#1}
    {\cBilat{\lattice}}
    {\cBilat{#1}}%
}

\newcommand{\Operator}{\m{O}}
\NewDocumentCommand\Op{g}{%
  \IfNoValueTF{#1}
    {\m{\Operator}}
    {\m{\Operator \left(#1\right)}}%
}

\newcommand{\Approximator}{\m{A}}
\NewDocumentCommand\Ap{g}{%
  \IfNoValueTF{#1}
    {\m{\Approximator}}
    {\m{\Approximator \left(#1\right)}}%
}

\newcommand{\calo}{\m{{I_o}}}

\NewDocumentCommand\struclat{g}{%
	\IfNoValueTF{#1}
		{\m{\lattice^\voc_\calo}}
		{\m{\lattice^\voc_{#1}}}
}
\NewDocumentCommand\domstruclat{g}{%
	\IfNoValueTF{#1}
		{\m{\lattice^\voc_\calo}}
		{\m{\lattice^\voc_{#1}}}
}

\NewDocumentCommand\proj{g}{%
	\IfNoValueTF{#1}
		{\m{p}}
		{\m{p(#1)}}
}

\newacronym{CNF}{CNF}{Causal Normal Form}

\keyword{naming}{naming}
\keyword{Naming}{Naming}
\keyword{Chfun}{$\D$-selection}
\keyword{chfun}{$\D$-selection}
\keyword{welldef}{well-defined}

\newcommand{\domainof}[1]{\m{D^{#1}}}

\newtheorem{thm}{Theorem}[section]

\newtheorem{definition}[thm]{Definition}

\newtheorem{example}[thm]{Example}

\newtheorem{ex*}{Example}

\newcommand{\asprule}{\m{\,\text{:-}\,}}

\RenewDocumentCommand\CLRule{m+g}{%
		\IfNoValueTF{#2}
			{\m{& #1.}}
			{\m{& #1 \cause #2 }}}

\lstdefinelanguage{br}{
		morekeywords=[1]{if,then,else,for,end,while,when,rule},
	}
	\newcommand{\Catom}[1]{#1}
	\NewDocumentCommand\Call{g+g+g}{%
	\IfNoValueTF{#1}%
	    {\m{\mathbf{All}}}%
	    {\m{\mathbf{All\,}#1[#2]: #3}}%
	 }

	\NewDocumentCommand\Cand{g+g}{%
	\IfNoValueTF{#1}%
	    {\m{\mathbf{And}}}%
	    {\m{#1\mathbf{\,And\,}#2}}%
	}
	\NewDocumentCommand\Csel{g+g+g+g}{%
	\IfNoValueTF{#1}%
	    {\m{\mathbf{Select}}}%
		{\IfNoValueTF{#4}%
			{\m{\mathbf{Select\,}#1[#2]: #3}}%
			{\m{\mathbf{Select}_{#1}\,#2[#3]: #4}}%
	}}
	\NewDocumentCommand\Cor{g+g+g+g+g}{%
	\IfNoValueTF{#1}%
		{\m{\mathbf{Or}}}%
		{\m{#1\mathbf{\,Or\,}#2}%
			\IfValueTF{#3}{\mathbf{\,Or\,}#3%
				\IfValueTF{#4}{\mathbf{\,Or\,}#4%
					\IfValueTF{#5}{\mathbf{\,Or\,}#5}{}%
				}{}%
			}{}%
		}%
	}
	\NewDocumentCommand\Cnew{g+g+g}{%
	  \IfNoValueTF{#1}%
	    {\m{\mathbf{New}}}%
	    {\IfNoValueTF{#3}%
			{\m{\mathbf{New\,}#1: #2}}%
			{\m{\mathbf{New\,}#1[#2]: #3}}}}%

	\NewDocumentCommand\Cif{g+g}{%
	\IfNoValueTF{#1}%
	    {rule}%
	    {\m{#2 \leftarrow #1 }}%
	}

\keyword{CEE}{CEE}

\renewcommand{\CLRule}[2]{\Cif{#2}{#1}}

\renewcommand{\subs}[2]{#2:#1}

\renewcommand{\struct}{\m{J}}

\newcommand{\tcaused}{endogenous\xspace}
\newcommand{\topen}{exogenous\xspace}

\newcommand{\xemph}[1]{\emph{#1}} 

\keyword{cset}{effect set} 
\keyword{poscset}{possible effect set} 

\setboolean{commentmargin}{false}

\renewcommand{\P}{\pred{P}}
\newcommand{\TB}{\pred{T}(\pred{B})}
\newcommand{\TS}{\pred{T}(\pred{S})}
\newcommand{\cheat}{}
\newcommand{\smallcheat}{}

\usepackage{supp}
\begin{document}
\title
{\foclog: A Knowledge Representation Language of Causality}

\author[Bart Bogaerts, et al.]
         {Bart Bogaerts, Joost Vennekens, Marc Denecker\\
            Department of Computer Science, KU Leuven\\
		E-Mail: firstname.lastname@cs.kuleuven.be
          \and Jan Van den Bussche\\
		Hasselt University \& transnational University of Limburg\\
		E-Mail: jan.vandenbussche@uhasselt.be}

\pagerange{\pageref{firstpage}--\pageref{lastpage}}
\volume{TODO} 
\jdate{TODO}
\setcounter{page}{1}
\pubyear{TODO}
\pubauthor{Bogaerts, et al.}
\jurl{TODO}
\pubdate{TODO}

\maketitle


\begin{abstract}
Cause-effect relations are an important part of human knowledge. 
In real life, humans often reason about complex causes linked to complex effects. 
By comparison, existing formalisms for representing knowledge about causal relations are quite limited in the kind of specifications of causes and effects they allow. 
In this paper, we present the new language \foidplus, which offers a significantly more expressive representation of effects, 
including such features as the creation of new objects. 
We show how \clog integrates with first-order logic, resulting in the language \foclog.
We also compare \foclog with several related languages and paradigms, including inductive definitions, disjunctive logic programming, business rules and extensions of Datalog.
\end{abstract}

	
\section{Introduction}\label{sec:intro} 
	

Cause-effect relations are an important part of human knowledge. 
There exist a number of knowledge representation languages \cite{McCainT96,journal/tplp/VennekensDB10,Cabalar12} 
in which logic programming style rules are used to represent such relations. 
The basic idea in all these approaches is that the head of such a rule represents an effect that is caused by its body. 
In this paper, we are particularly concerned with \cpl \cite{journal/tplp/VennekensDB10}. 
More specifically, we consider the variant of \cpl without
probabilities, and we will extend this language with three
features: 
dynamic non-determinsitic
choice; object creation; and  
recursive nesting of cause-effect relations. 
We call the resulting language
\foidplus.  
In this paper, we present \clog and its informal semantics. 
For the formal semantics, we refer to an accompanying technical report \cite{cw656/BogaertsVDV14}. 
We also present the integration of \clog with first-order logic, and thus show that \clog fits in the \fodot Knowledge Base System project \cite{ASPOCP/Denecker12}.

 Let us begin
by recalling the guiding principles behind \cpl.
When compared to predecessors, such as the causal logic of \citeN{McCainT96}, 
one of the important contributions of this languages is to add two modelling principles that are common in causal modelling. 
The first is the distinction between {\em endogenous} and {\em exogenous} properties, 
i.e., those whose value is determined by the causal laws in the model and those whose value is not, respectively \cite{Pearl00}. 
The second is the {\em default-deviant} assumption, used also by, e.g., \citeN{Hall04} and \citeN{Hitchcock07}. 
The idea here is to 
assume that each endogenous property of the domain has some ``natural'' state, that it will be in whenever nothing is acting upon it. 
For ease of notation, \cpl identifies the default state with falsity, and the deviant state with truth.
For example, consider the following simplified model of a bicycle, in which a pair of gear wheels can be put in motion by pedalling:
\begin{align}
\pred{Turn}(\pred{BigGear})   &\leftarrow \pred{Pedal}.         \label{pedal} \\
\pred{Turn}(\pred{BigGear})   &\leftarrow \pred{Turn}(\pred{SmallGear}).\label{bigg}   \\
\pred{Turn}(\pred{SmallGear}) &\leftarrow \pred{Turn}(\pred{BigGear}).  \label{smallg}
\end{align}
Here, $\pred{Pedal}$ is exogenous, while $\pred{Turn}(\pred{BigGear})$ and $\pred{Turn}(\pred{SmallGear})$ are endogenous. The semantics of this causal model is given by a straightforward ``execution'' of the rules. The domain starts out in an initial state, in which all endogenous atoms have their default value {\em false} and the exogenous atom $\pred{Pedal}$ has some fixed value. If $\pred{Pedal}$ is true, then the first rule is applicable and may be fired (``$\pred{Pedal}$ causes $\pred{Turn}(\pred{BigGear})$'') to produce a new state of the domain in which $\pred{Turn}(\pred{BigGear})$ now has its deviant value {\em true}. In this way, we construct the following sequence of states (we abbreviate symbols by their first letter):
\begin{equation}\label{branch}
\smallcheat \{\P\} \cheat \overset{\eqref{pedal}}{\rightarrow} \cheat\{\P,\TB\} \cheat\overset{\eqref{smallg}}{\rightarrow}\cheat \{\P,\TB,\TS\} \cheat\overset{\eqref{bigg}}{\rightarrow} \cheat\{\P,\TB,\TS\}\cheat \cheat  \smallcheat
\end{equation}
Note that firing rule \eqref{bigg} does not change the state of the world, because its effect is already true. Moreover, it is obvious that this will always be the case, so this rule may seem redundant. However, many interesting applications of causal models require the use of interventions \cite{Pearl00}, e.g., to evaluate counterfactuals or to predict the effects of actions. 
As shown by \citeN{journals/lncs/VennekensDB10}, rule \eqref{bigg} allows \cpl to represent this example in a way that produces the correct results for all conceivable interventions in a manner that is more modular and more concise than, among others, Pearl's structural models \cite{Pearl00}.

After rules \eqref{pedal}, \eqref{smallg} and \eqref{bigg} have all fired, there are no more rules left whose body is satisfied and that have not yet fired. At this point, the process is at an end and the domain has reached a final state. It is this final state, rather than the details of the intermediate process, that we are really interested in. One of the most important properties of \cpl is that, while there may be any number of different processes derived from a causal theory, the final state that is eventually reached is unique for any given interpretation for the exogenous predicates---at least, for examples such as this one. 
In general, \cpl also allows rules with a {\em non-deterministic} effect, such as:
\[
\Cor{(\pred{Turn}(\pred{SmallGear}):0.99)}{ (\pred{ChainBreaks}:0.01)} \leftarrow \pred{Turn}(\pred{BigGear}).
\]
Now, the cause $\pred{Turn}(\pred{BigGear})$ produces one of two possible effects, and there is an associated probability distribution over these two possibilities. The effect on the semantics is that, instead of a linear progression of states as in \eqref{branch}, we get a tree structure in which each firing of a non-deterministic rule introduces a branching of possibilities. 
 When considering also the probabilities associated to non-deterministic choices, the tree \ignore{becomes a probability tree, which }defines a probability distribution over its leaves, i.e., over the final states that  may be reached. 
It was shown in \cite{journal/tplp/VennekensDB10} that, given a specific interpretation for the exogenous atoms, this 
distribution is unique, even 
though there may exists many probability trees that produce it. 

In many circumstances, the precise values of the probabilities are not of interest. In such cases, a non-probabilistic variant of \cpl may be used, in which these are omitted.  The head of a rule is then simply a disjunction:
\[ \Cor{\pred{Turn}(\pred{SmallGear})}{ \pred{ChainBreaks} }
\leftarrow \pred{Turn}(\pred{BigGear}).
\]
The trees then no longer produce a probability distribution over final states, but they describe the set of all final states that may be reached. 
In other words, this formalism has a possible world semantics. It is this non-probabilistic variant that concerns us in this paper.

Like other rule-based approaches to causality, \cpl uses a very simple way of specifying the possible effects of some cause, namely, as a disjunction of ground atoms. Clearly, this does not---or, at least, not directly---cover many interesting phenomena that may occur in practice:
\begin{itemize}
\item A robot enters a room, opens some of the doors in this room, and then leaves by one of the doors that are open. The robot's leaving corresponds to a non-deterministic choice between a {\em dynamic} set of alternatives, which is determined by the robot's own actions, and therefore cannot be hard-coded into the head of a rule. A language construct for representing such choices is present in P-log \cite{BaralGR04}. 
\item A stallion and a mare that are put in the same field may cause the birth of a foal. Therefore, not only the properties of these horses are governed by causal laws, but also their very  existence. 
\item A horse being the parent of a foal is itself a cause for its own height to have a causal link to the height of the foal. Therefore, causal laws may be nested, in the sense that an effect can itself again consist of an entire causal law.   
\end{itemize}
The goal of this paper is to develop an expressive knowledge representation language that is able to represent these more complex effects, and others like them, in a direct way. Moreover, we want to do this in a way that extends the approach of \cpl. To summarise, the formal semantics of the language should consist of a set of possible worlds, each of which can be constructed by a non-deterministic causal process. This process will take place in the context of a fixed interpretation for the exogenous atoms. It will start from an initial state in which each of the endogenous atoms is at its default value false. The causal laws of our language will then ``fire'' and flip atoms 
to their deviant value, until no more such flips are possible. Whereas in \cpl these flips happen one atom at a time, our extended language will flip {\em sets} of atoms at the same time.
Moreover, our logic will present syntax and semantics for object-creation, as is needed in the second of the above examples.

The rest of this paper is structured as follows: we start by introducing causal effect expressions (\CEEs) and their informal semantics in Section \ref{sec:syntax}. 
In Section \ref{sec:foclog}, we explain how \foidplus is integrated with first-order logic, resulting in \foclog, a member of the \fodot family of extensions of first-order logic.
We conclude  in Section \ref{sec:comparison} by comparing \foidplus with various other paradigms, including inductive definitions \cite{tocl/DeneckerT08}, disjunctive logic programming with existential quantifications  \cite{tplp/YouZZ13},  Business Rules systems \cite{Group2000Defining} and Datalog extensions \cite{datalog/GreenAK12}.

\section{Syntax and Informal Semantics}\label{sec:syntax} 									
	We assume familiarity with basic concepts of \FO. 
Vocabularies, formulas, and terms are defined as usual. 
A \voc-structure \I interprets all symbols (including variable symbols) in a vocabulary \voc;  \domainof{\I} denotes the domain of \I and
$\sigma^\I$, with $\sigma$ a symbol in $\voc$, denotes
the interpretation of $\sigma $ in \I.
We use $\I[\subs{v}{\sigma}]$ for the structure $\J$ that equals \I, except on $\sigma$: $\sigma^\J=v$.
\emph{Domain atoms} are atoms of the form $P(\ddd)$ where the $d_i$ are domain elements.
We use restricted quantifications \cite{PreyerP02}, e.g., in \FO, these are formulas of the form
$\univBQ{x}{\psi}: \varphi\label{formula:BinQuantForall}$ or $\exisBQ{x}{\psi}: \varphi\label{formula:BinQuantExists}$,
meaning that $\varphi$ holds for all (resp.~for a) $x$ such that $\psi$ holds.
The above expressions are syntactic sugar for
$	\forall x: \psi \limplies \varphi
$ and $ 
	\exists x: \psi \land \varphi,
$
but such a reduction is not possible for other restricted quantifiers that we will define below.
We call $\psi$ the \emph{qualification} and $\varphi$ the \emph{assertion} of the restricted quantifications.
From now on, let \voc be a relational vocabulary, i.e., \voc consists only of predicate, constant and variable symbols. 

\subsection{Syntax}

\begin{definition}
 \emph{Causal effect expressions} (\CEE) are defined inductively as follows:
\begin{itemize}
	\item if $P(\ttt)$ is an atom, then $\Catom{P(\ttt)}$ is a \CEE,
	\item if $\varphi$ is a first-order formula and $C'$ is a \CEE, then $\Cif{\varphi}{C'}$ is a \CEE,
	\item if $C_1$ and $C_2$ are \CEEs, then $\Cand{C_1}{C_2}$ is a \CEE,
	\item if $C_1$ and $C_2$ are \CEEs, then $\Cor{C_1}{C_2}$  is a \CEE,
	\item if $x$ is a variable, $\varphi$ is an \fo formula and $C'$ is a \CEE, then $\Call{x}{\varphi}{C'}$  is a \CEE,
	\item if $x$ is a variable, $\varphi$ an \fo  formula and $C'$ a \CEE, then $\Csel{x}{\varphi}{C'}$ is a \CEE,
	\item if $x$ is a variable and $C'$ is a \CEE, then $\Cnew{x}{C'}$ is a \CEE.
\end{itemize}
\end{definition}

We call a \CEE an \emph{atom-expression}  (respectively \emph{\Cif-},   \emph{\Cand-}, \emph{\Cor-},  
\emph{\Call-}, \emph{\Csel-} or \emph{\Cnew-expression}) if it is of the corresponding form. 
We call a predicate symbol $P$ \xemph{\tcaused}in $C$ if $P$ occurs as the symbol of a (possibly nested) atom-expression in $C$, i.e., 
if $P$ occurs in $C$ but not only in first-order formulas, i.e., not only in qualifications of restricted \clog quantifications (\Call and \Csel) or conditions of \Cif-expressions.
All other symbols are called \xemph{\topen}in $C$. This is a straightforward generalisation of the same notions in \cpl.
An occurrence of a variable $x$ is \emph{bound} in a \CEE if it occurs in the scope of a quantification over that variable 
($\forall x$, $\exists x$, $\Call\, x$, $\Csel\, x$, or $\Cnew\, x$) and \emph{free} otherwise. A variable is \emph{free} in a \CEE if it has free occurrences. 
A \emph{causal theory}, or \emph{\foidplus theory} is a \CEE without free variables. 
By abuse of notation, we often represent a causal theory as a set of \CEEs; the intended causal theory is the \Cand-conjunction of these \CEEs.
We often use \D for a causal theory and $C$, $C'$, $C_1$ and $C_2$ for its subexpressions. 


\subsection{Informal Semantics of \CEEs}
We now present the informal semantics of \CEEs, due to space restrictions, the formalisation of this semantics is lacking in this paper. 
For a complete description of the formal semantics, we refer to an accompanying technical report \cite{cw656/BogaertsVDV14}.
A \CEE is a description of a set of causal laws.
In the context of a state of affairs---which we represent, as usual, by a structure---a \CEE non-deterministically describes a 
set of effects, i.e., a set of events that take place and change the state of affairs.
We call such a  set the \xemph{\cset}of the \CEE.
From a \CEE $C$, we can derive causal processes similar to \eqref{branch}; a causal process is a sequence of intermediate states, starting from the default state, such that, at each state, the effects described by $C$ take place. 
The process ends if the effects no longer cause changes to the state.
A structure is a model of a \CEE if it is the final result of such a process.
We now explain in a compositional way what the \cset  of a \CEE is in a given state of affairs.

The effect of an atom-expression $A$ is that $A$ is flipped to its deviant state.
A conditional effect, i.e., a rule expression, causes the \cset  of its head if its body is satisfied in the current state, and   nothing otherwise.
The \cset described by an \Cand-expression is the union of the \csets of its two subexpressions;  an \Call-expression $\Call{x}{\varphi}{C'}$ causes 
the union of all \csets of $C'(x)$ for those $x$'s that satisfy $\varphi$.
An expression $\Cor{C_1}{C_2}$ non-deterministically causes either the \cset of $C_1$ or the \cset of $C_2$; a \Csel-expression $\Csel{x}{\varphi}{C'}$ causes the \cset 
of $C'$ for a non-deterministically chosen   $x$ that satisfies $\varphi$. 
An object-creating \CEE $\Cnew{x}{C'}$ causes the creation of a new domain element $n$ and  the \cset of $C'(n)$.

\begin{example}\label{ex:american}
	Permanent residence in the United States can be obtained in several ways. 
	One way is  passing the naturalisation test.
	Another way is by playing the ``Green Card Lottery'', where each year a number of lucky winners are  randomly selected and granted permanent residence.
%
	We model this as follows:
	\begin{align*}
		&\Call{p}{\pred{Applied}(p) \land \pred{PassedTest}(p)}{ \pred{PermRes}(p)}
		\\ &
	\CLRule{(\Csel{p}{\pred{Applied}(p)}{\pred{PermRes}(p)})}{\pred{Lottery}.}
	\end{align*}
	The first \CEE describes the ``normal'' way to obtain permanent residence; 
	the second rule expresses that one winner is selected among everyone who applies. 
%
If \I is a structure in which $\pred{Lottery}$ holds, due to the non-determinism, there are many \poscsets of the above \CEE, namely all sets $\{\pred{PermRes}(p) \mid p\in \domainof{\I} \land \inter{ p\in\pred{Applied}}{\I} \land \inter{\pred{PassedTest}(p)}{\I} \}$ $ \cup$ $\{\pred{PermRes}(d)\}$ for some  $d\in\inter{\pred{Applied}}{\I}$. 
The two  \CEEs are considered independent: the winner could be one of the people that obtained it through standard application, as well as someone else.
Note that in the above, there is a great asymmetry between $\pred{Applied}(p)$, which occurs as a qualification of \Csel-expression, 
and  $\pred{PermRes}(p)$, which occurs as a caused atom, in the sense that the effect will never cause atoms of the form $\pred{Applied}(p)$, but only atoms of the form $\pred{PermRes}(p)$. 
This is one of the cases where the qualification of an expression cannot simply be eliminated. 
\end{example}

%

\begin{example}\label{ex:mail}
Hitting the ``send'' button in your mail application causes the creation of a new package containing a specific mail. 
That package is put on a channel and will be received some (unknown) time later. 
As long as the package is not received, it stays on the channel.
In \foidplus, we model this as follows:
\begin{align*}
	&\Call{m,t}{\pred{Mail}(m)\land \pred{HitSend}(m,t)}{\Cnew{p}{  
		\Cand{\pred{Pack}(p)}{\Cand{\pred{Cont}(p,m)}{\Cand{\pred{OnCh}(p,t+1)\\  &\quad }{
		\Csel{d}{d>0}{\pred{Received}(p,t+d)}                 }}}              }          }\\
	&\Call{p, t}{\pred{Pack}(p) \land \pred{OnCh}(p,t) \land \lnot \pred{Received}(p,t)}{
	\pred{OnCh}(p,t+1)}
	\end{align*}
	
	Suppose an interpretation $\inter{\pred{HitSend}}{\I}=\{(\pred{MyMail},0)\}$ is given.  
	A causal process then unfolds as follows: it starts in the initial state, where all \tcaused predicates are false.
	The \cset of the above causal effect in that state consists of 1) the creation of one new domain element, say $\_p$, and 2) the caused atoms $\pred{Pack}(\_p)$, $\pred{Cont}(\_p,\pred{MyMail})$, $\pred{OnCh}(\_p,1)$ and $\pred{Received}(\_p,7)$, 
	where instead of $7$, we could have chosen any number greater than zero. 
	Next, it continues, and in every step $t$, before receiving the package, an extra atom $\pred{OnCh}(p,t+1)$ is caused. 
	Finally, in the seventh step, no more atoms are caused; the causal process ends. 
	The final state is a model of the causal theory.

%
%

\end{example}

\section{\foclog: Integrating FO and \foidplus}\label{sec:foclog}
	First-order logic  and  \foidplus have a straightforward integration, \foclog. Theories in this logic are sets of FO sentences and \CEEs. A model of such a theory is a structure that satisfies each of its expressions (each of its \CEEs and formulas).  An illustration is  the mail protocol from  Example~\ref{ex:mail}, which we can extend with the ``observation''  that at at some time point, two packages are on the channel:
$ \exists t, p_1, p_2: \pred{OnCh}(p_1,t) \land \pred{OnCh}(p_2,t) \land
 p_1\neq p_2.$
 Models of this theory represent states of
affairs where at least once two packages are on the channel simultaneously.  This entirely differs from \Cand-conjoining our \CEE with
 \begin{align*}
 \Csel{t, p_1, p_2}{ p_1 \neq p_2}{}
	\Cand{\pred{OnCh}(p_1,t)}{ \pred{OnCh}(p_2,t)}.
 \end{align*}
The resulting \CEE would have unintended models in which  two  packages suddenly appear on the channel for no reason. 
\renewcommand{\arraystretch}{1.3}

In \foclog,  \Cnew-expressions can be simulated with \Csel-expressions together with \FO axioms expressing the unicity of the newly ``created'' objects. E.g., 
    \[\Cand{\Cnew{x}{P(x,a)}}{\Cnew{x}{Q(x)}}\] 
is simulated by introducing auxiliary unary predicates $N_1$ and $N_2$ that identify the objects created by the expressions and writing:
\begin{align*}
&\left\{\Cand{  (\Csel{x}{\Tr}{(\Cand{N_1(x)}{P(x,a)})})}{
\Csel{x}{\Tr}{(\Cand{N_2(x)}{Q(x)})} }\right\}\\
 & \forall x: \neg (N_1(x)\land  N_2(x))
\end{align*}
It is clear that \Cnew-expressions are more natural and more modular than this simulation. 

Despite the syntactical correspondence between \CEEs and \FO formulas
($\Cand$ corresponds to $\land$, $\Call$ to $\forall$, \dots), it is
obvious that they have an entirely different meaning, and that both are
useful. This is why we chose to introduce new connectives rather than
overloading the ones of \FO.  The logic \foclog has further
interesting extensions, e.g., by adding aggregates in \FO formulas,
including in qualifications and conditions of \CEEs.
\section{Comparison and Future Work}\label{sec:comparison} 								
	In this section, we compare \foclog to other existing paradigms. This comparison is only an initial study. By the time of publishing, a more extended version of a comparison between \foclog and other paradigms has appeared \mycite{examplesClog}.

Due to its simple recursive syntax, \foclog is a very general logic
that generalises several existing logics and shows overlaps with many
others in different areas of computational logic.  \foidplus is an
extension of (the non-probabilistic version of) \cpl. \foclog is
an extension of the logic \foid~\cite{tocl/DeneckerT08}. An \foid theory is a set of \fo sentences
and inductive definitions (ID), which are sets of rules of the form
\[\forall \xxx: P(\ttt)\lrule \varphi,\] where $\varphi$ is an \fo
formula. Such a rule corresponds to a \CEE \[\Call{\xxx}{\varphi}{P(\ttt)} \] or equivalently,
\[\Call{\xxx}{\Tr}{(\Cif{\varphi}{P(\ttt)})}\] and a definition corresponds to the \Cand-conjunction of its rules.
The semantics of \foid corresponds exactly to the semantics of the corresponding \foclog theory~\cite{cw656/BogaertsVDV14}.
\citeN{DeneckerTDB98}  already pointed to the correspondence between
causality and inductive definitions and exploited it for  solving the
causal {\em ramification problem} of temporal reasoning
\cite{McCHay69}. The \CEEs presented here can be seen as a
non-deterministic extension of inductive definitions with an informal semantics based on
causal processes.

\foclog  shows similarity to extensions of disjunctive logic programming (DLP) such as DLP  with existential quantification in rule heads 
\cite{tplp/YouZZ13} and the  stable semantics for \FO as defined by \citeN{AI/FerrarrisLL11}.
Here constraints correspond to \fo sentences in \foclog and other rules correspond to \clog expressions.
However, there is an important semantical difference. Suppose we want to express  Example \ref{ex:american}, where all people passing a test and one random person are given permanent residence in the United States.
The E-disjunctive program
  \begin{align*}
  &\exists X: permres(X)\asprule lottery\\
   &\forall X: permres(X) \asprule passtest(X)
 \end{align*}
is similar to  \begin{align*} \Cand{\Cif{lottery}{\left(\Csel{x}{\Tr}{permres(x)\right)}}\\}{\Call{x}{passtest(x)}{permres(x)}}
 \end{align*}
Semantically, the E-disjunctive program imposes a minimality condition: the lottery is always won by a person succeeding the test, if there exists one.  On the other hand, in \foclog the two rules execute independently, and models might not be minimal. In this example, it is the latter that is intended. We believe that one advantage of \foidplus is its clear causal informal semantics. On the other hand, there are ways to simulate the causal semantics and  the \Cnew operator of \foidplus in E-disjunctive programs while it follows from complexity arguments that  not all E-disjunctive programs can be expressed in \foclog~\mycite{inferenceClog}.

 Other semantics than the stable semantics for DLP have been developed. 
 For example, \citeN{BrassD96} defined D-WFS, a well-founded semantics  for DLP.
 This semantics has the property that if a program contains two identical lines, one of them can be removed. 
 However, in our context, a duplicate effect means that a same causal effect happens twice (maybe for different reasons), independently, and hence different choices
 might be made in each of these rules. 

 The logic of cause and change \cite{McCainT96} differs from  \foidplus in several important aspects; 
 in McCain \& Turner's logic  both true and false atoms need a cause. 
 In \foidplus on the other hand,  \tcaused predicates can be false (the default value) 
 without reason but can only be true (the deviant value) if caused. 
 Moreover, we rule out unfounded ``cyclic'' causation. 
 For instance, if $\pred{Pedal}$ is false, in \foidplus, $\pred{Turn}(\pred{BigGear})$ and $\pred{Turn}(\pred{SmallGear})$ 
 are false but in McCain and Turner's logic they may be true and caused by each other.  
 We call this ``spontaneous generation'' and do not admit it in \foidplus.
\ignore{  For instance, in  the gearwheel example in the introduction athe  $\Cif{C}{C}$ has a unique model in which $C$ is false. In McCain \& Turner's language on the other hand, a similar causal theory has a unique model in which $C$ holds. 
In practice,
the main advantage of McCain \& Turner’s treatment of causal cycles seems to be that it offers a way to introduce \topen atoms into the language. Indeed, by
including both $Q \lrule Q$ and $\lnot Q \lrule \lnot Q$, one  expresses that $Q$ can have any truth
value. Of course, \foidplus
has no need for such a mechanism, since we make an explicit distinction between
\topen and \tcaused predicates.
Another advantage of \foidplus is its expressive recursive syntax.}

We find operators similar to those of \foidplus in several other formalisms. For example, \Csel-, \Call-, \Cor- and \Cif-expressions  are present in  the subformalism of the language Event-B that serves to specify effects of actions \cite{BookAbrial2010}.  The \Cnew operator  is found in various other rule based paradigms, for example in 
Business Rules systems \cite{Group2000Defining}.
The JBoss manual \cite{browne2009jboss} contains the following rule: 
 \begin{lstlisting}
 when Order( customer == null ) then  insertLogical(new ValidationResult(“validation.customer.missing”));\end{lstlisting}
meaning that if an order is created without customer, a new \textit{ValidationResult} is created with the message that the customer is missing. This can be translated to \foidplus as follows:
\begin{align*}
	&\Call{y}{\pred{Order}(y)\land\pred{NoCustumer}(y)}{
	\Cnew{x}{\Cand{\pred{ValidationR}(x)}{\pred{Message}(x,\text{``\dots''})}}}.
\end{align*}

Another field in which related language constructions  have been developed is the field of deductive databases. 
\citeN{jcss/AbiteboulV91} considered various extensions of Datalog, resulting in non-deterministic semantics for queries and updates. 
One of the studied extensions is object creation. 
Such an extension is present in the LogicBlox system \cite{datalog/GreenAK12}.  An example from the latter paper is the rule: 
$ President(p), presidentOf[c] = p \lrule Country(c)$
which means that for every country $c$, a new (anonymous) ``derived
entity'' of type $President$ is created. Of course, this president is not a new person, but it is new with respect to a given database. 
Such rules with implicit existentially
quantified head variables correspond with $\Cnew$-expressions in
\foidplus. 

Other Datalog extensions with other
forms of object creation exist. For example \citeN{VandenBusscheP95} discuss a
version with creation of sets and compare its expressivity 
with simple object creation. 

Non-deterministic choices have been studied intensively in the context
of deductive databases.  \citeN{conf/jcdkb/KrishnamurthyN88} introduced a
non-deterministic choice in Datalog.  This choice was
{\it static}: choice models are constructed in three steps.  First,
models are calculated while ignoring choices (choosing everything);
second, this model is used to select a number of choices for all
occurrences of {\it choice goals} and third, models are recalculated
with respect to these choice goals.  In other work
\cite{pods/SaccaZ90,GianottiPSZ91}, it is argued that static choices
do not behave well in the presence of recursion; hence {\it dynamic}
choices were introduced.  \citeN{pods/SaccaZ90} use stable models  to provide a model-theoretic description of these dynamic
choices.  \citeN{jlp/WeidongJ96} introduced an alternative choice principle on
predicates $P$. There, the values in certain argument
positions in the tuples of $P$ are chosen non-deterministically in
function of the values at the other argument positions.  The semantics of
that logic is based on the well-founded semantics; this
choice principle is very different from the principle in \foidplus. Compared to
these, \foidplus resembles most the language of \citeN{pods/SaccaZ90};
the difference is that \foidplus supports a recursive syntax and is
based on the well-founded semantics, whereas \citeN{pods/SaccaZ90} use
stable semantics.



The above similarities suggest that \foclog is a promising language
to study and unify many existing logical paradigms and to provide a clear
informal semantics for them.
An in-depth semantical analysis of the exact 
relationship between \foclog and the languages described above is an
interesting  topic for future work.
Another research challenge is  extending \foclog with types, function symbols, arithmetic, etc. in order to make it useful as a KR-language.  
We need to study the complexity of various inference tasks in  \foclog, and develop and implement  algorithms for these  various  tasks. 
By the time of publication, a first study of complexity and inference in \foclog has appeared \mycite{inferenceClog}.
Another research question is to add probabilities to \foidplus  to obtain an extension of the probabilistic  \cpl, and possibly also of other related logics such as BLOG \cite{ijcai/MilchMRSOK05} and P-Log  \cite{BaralGR04}. 





\bibliography{krrlib}

\bibliographystyle{acmtrans}
\end{document}

%
